\documentclass[smallextended]{svjour3}

\usepackage{fullpage}

\usepackage{graphicx}

\usepackage{hep}

\title{Exponential quantum enhancement for distributed addition with local nonlinearity}

\author{Adam Henry Marblestone \and Michel Devoret}

\institute{
	Adam Henry Marblestone
	\and
	Michel Devoret \at
	Departments of Applied Physics and Physics, Yale University
}

\date{2009-07-20}

\begin{document}

\maketitle

\begin{abstract}
We consider classical and entanglement-assisted versions of a distributed computation scheme that computes nonlinear Boolean functions of a set of input bits supplied by separated parties. Communication between the parties is restricted to take place through a specific apparatus which enforces the constraints that all nonlinear, nonlocal classical logic is performed by a single receiver, and that all communication occurs through a limited number of one-bit channels. In the entanglement-assisted version, the number of channels required to compute a Boolean function of fixed nonlinearity can become exponentially smaller than in the classical version. We demonstrate this exponential enhancement for the problem of distributed integer addition.

\end{abstract}

\tableofcontents

\section{Introduction}

The study of quantum communication complexity \cite{QuantumCommunicationComplexityASurvey} \cite{MultipartyQuantumCommunicationComplexity} \cite{NonlocalCorrelationsAsAnInformationTheoreticResource} has drawn attention to the feasibility using quantum protocols of certain classically impossible communication tasks. This is strikingly demonstrated by quantum pseudo-telepathy protocols \cite{QuantumPseudoTelepathy}. In quantum pseudo-telepathy, entanglement eliminates the classical need for signaling between separated parties collaborating to perform a task. Crucially, while entanglement \emph{substitutes} for signaling in pseudo-telepathy scenarios, it does not itself \emph{produce} superluminal signaling \cite{QuantumInformationAndRelativityTheory}. Rather, entanglement and communication are partially interchangeable resources. Here we demonstrate a new exponential quantum-classical dichotomy in multi-party communication, in which entanglement is partially interchangeable with nonlocal, nonlinear logic. We consider classical and entanglement-assisted versions of a distributed computation scheme that allows a single receiver to compute nonlinear Boolean functions of a set of input bits supplied by spatially separated senders. Communication between all parties is restricted to take place through a specific apparatus which enforces the constraints that all nonlinear, nonlocal classical logic is performed by the single receiver, and that the senders communicate only with the receiver through a limited number of linear one-bit channels. In the entanglement-assisted version, where local rotations on entangled qubits are performed and where the results of local measurements on these qubits can be sent through the linear communication apparatus, the number of channels required to compute a Boolean function of fixed nonlinearity can become exponentially smaller than in the classical version. Our imposed handicap on the communication apparatus, which limits the amount of communication and prevents the communication process itself from introducing nonlinearity into the Boolean function being distributedly evaluated, reveals a key difference between the quantum and classical versions of the system: in the quantum version, the nonlinearity index of the computable functions can grow exponentially with the number of channels, whereas in the classical version it grows linearly. We use the entanglement-assisted version to demonstrate distributed integer addition in this setup using exponentially fewer channels than would classically be necessary.

Although the use of a communication handicap to demonstrate this quantum-classical dichotomy is in a sense artificial, the specific communication setup used, which we call \emph{locally nonlinear distributed evaluation}, is quite natural for the purpose of drawing attention to the distinction between linearity and nonlinearity in the communication process. In particular, it presents a fixed number of linear combinations of the senders' input bits to a receiver capable of executing arbitrary classical nonlinear logic on bits locally accessible to her, while preventing the senders and the receiver from performing any extraneous communication or nonlocal nonlinear logic. Using this setup, nonlinearity in the communication process can be precisely quantified in terms of the required number of one-bit channels. For distributed integer addition, the required number of channels classically grows as the number of senders, whereas in the quantum scenario it grows as the logarithm of the number of senders. Our results should also apply to any other communication setup that imposes the same constraints, i.e., the setup must present the receiver with a fixed number of linear combinations of the senders' input bits, while preventing any further communication between the parties.

\section{Linear and nonlinear Boolean functions}

For Boolean functions $y(x_0,x_1,\cdots,x_{k-1})$ of $k$ Boolean variables, linearity is defined to be the condition that there exist Boolean constants $c_i$ and $b$ such that \begin{equation} \label{y from x one output bit} y(x_0,x_1,\cdots,x_{k-1}) = c_0 \cdot x_0 \oplus c_1 \cdot x_1 \oplus \cdots \oplus c_{k-1} \cdot x_{k-1} \oplus b \end{equation} where $\cdot$ denotes \textbf{AND} (product modulo $2$) and $\oplus$ denotes \textbf{XOR} (sum modulo $2$). Here linearity and affinity are treated as synonymous. On the other hand, an arbitrary Boolean function $f$ of $k$ bits is a polynomial of degree $\leq k$ in its inputs, with coefficients in the set $\{0,1\}$, addition given by $\oplus$ (\textbf{XOR}) and multiplication given by \textbf{AND}. \begin{eqnarray}
f\left(x_0, \cdots, x_{k-1} \right) = \nonumber \\ 
\left[A \oplus A_0 \cdot x_0 \oplus A_1 \cdot x_1 \oplus \cdots \oplus A_{k-1} \cdot x_{k-1} \right] \nonumber \\ 
\oplus \nonumber \\
\left( A_{01} \cdot x_0 \cdot x_1 \oplus A_{02} \cdot x_0 \cdot x_2 \oplus \cdots \oplus A_{k-1k-2} \cdot x_{k-1} \cdot x_{k-2} \right) \label{algebraic normal form}\\
\oplus \nonumber \\
\cdots \nonumber \\
\oplus \nonumber \\
\left( A_{0 1 2 \cdots k-1} \cdot x_0 \cdot x_1 \cdots x_{k-1} \right) \nonumber
\end{eqnarray} In this expression, brackets are placed around the linear part of the general function, while the nonlinear parts are put in parenthesis. Note that some of the coefficients $A_{ijk...}$ may be zero. This way of writing an arbitrary Boolean function as a polynomial is known as the \emph{algebraic normal form}, or \emph{Zhegalkin polynomial}. A crucial concept is the \emph{order of nonlinearity} of a Boolean function. The order of nonlinearity is sometimes called the \emph{algebraic degree}.

\begin{definition} \textbf{Order of nonlinearity $O(f)$ of a Boolean function $f$}: the order of nonlinearity $O(f)$ of a Boolean function $f$ is the degree of its Zhegalkin polynomial. In other words, $O(f)$ is the maximum number of input bits combined in a multi-way \textbf{AND} statement with \emph{non-zero} coefficient in the algebraic normal form representation \eqref{algebraic normal form} of $f$.\end{definition}

For example, the function $f(a,b,c, d) = b \cdot d \oplus a \cdot c \oplus a \cdot b \cdot d$ has $O(f) = 3$. It will be of no surprise that Boolean functions with orders of nonlinearity $\leq 1$ are called \emph{linear} ($O(f) = 0$ implies that $f$ is a constant), and that Boolean functions with orders of nonlinearity $>1$ are called \emph{nonlinear}. Nonlinear Boolean functions have algebraic normal form representations that contain at least a two-way \textbf{AND} statement. We say that a function $f$ with order of nonlinearity $O(f) = n$ is \emph{$n$-nonlinear}. Nonlinear Boolean functions are essential for computation, since any family of classical logic gates generating a Turing universal circuit model of computation must contain at least one nonlinear gate, such as \textbf{AND} or \textbf{OR} ($x \mathbf{OR} y = \overline{\bar{x}\mathbf{AND}\bar{y}} = x \oplus y \oplus x \cdot y$).

\section{Locally nonlinear distributed evaluation}

In the sections that follow, we consider a distributed computation scheme with the following properties, which define the notion of a \emph{locally nonlinear distributed evaluation}. We call the scheme \emph{locally nonlinear} because all nonlinear classical logic is performed locally, by a single receiver. 

\begin{definition} \textbf{Locally nonlinear distributed evaluation}
	
	\begin{figure}
		\begin{center}
	  \includegraphics[width = 450pt]{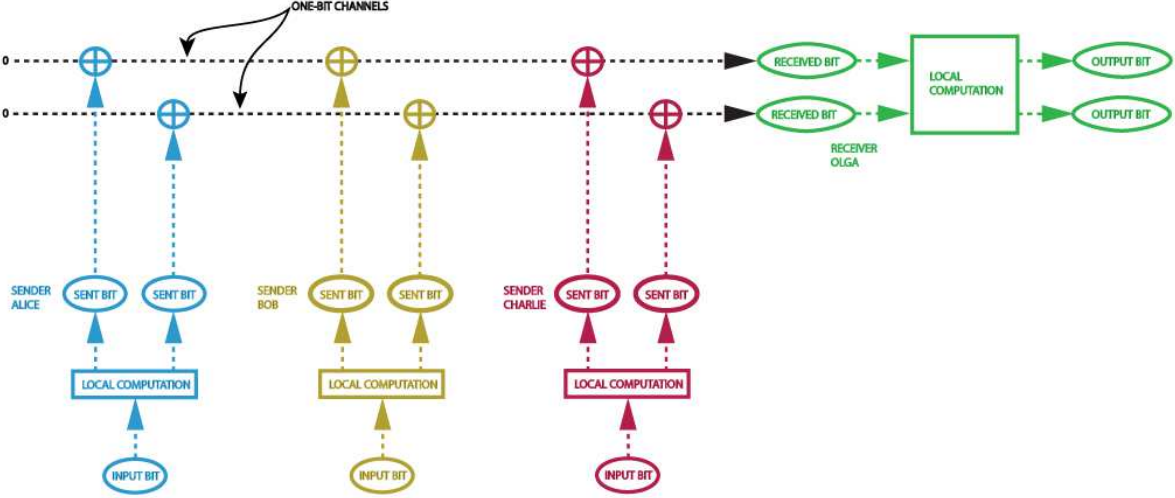}
	\end{center}
	\vspace{15pt}

	\caption{\textbf{Locally nonlinear distributed evaluation uses a linear, unidirectional communication bus with a fixed number of one-bit channels.} In this example, there are three senders (Alice, Bob and Charlie) and two channels. Each channel can carry only one bit to the receiver Olga. The symbol $\oplus$ indicates a classical \textbf{CNOT}. Note the appearance of the four types of bits in this setup: input bits, sent bits, received bits and output bits.}
	\label{fig:Figure-1}
	\end{figure}
	
\begin{enumerate} \item \textbf{Senders:} There are $N$ \emph{senders} (Alice, Bob, Charlie, ...), each possessing one bit from a collection of $N$ bits \[ x_0, x_1, \ldots, x_{N-1} \] The senders' possessed bits $x_0, x_1, \ldots, x_{N-1}$ are called \emph{input bits} for short. \item \textbf{Receiver:} There is one \emph{receiver}, Olga. \item \textbf{Linear, unidirectional communication bus:} As shown in \textbf{Figure \ref{fig:Figure-1}}, each sender may upload one single bit (vertical lines) to each of a fixed number $m$ of separate single-bit channels (horizontal lines) which reach Olga. The uploaded bits are called \emph{sent bits} to distinguish them from the senders' input bits. The $m$ bits that Olga receives from the channels are called \emph{received bits} to distinguish them from \emph{sent bits} and \emph{input bits}. The uploading process is linear, relying only on the \textbf{XOR} (i.e., \textbf{CNOT}) operation: the channels are initialized to zero, and if the value of a channel is $v$ before the $i^{th}$ sender uploads a bit $a_i$ to it, then the value of the channel after the $i^{th}$ sender's upload is $v \oplus a_i$. The value that Olga receives from the channel is then $\sum_{\oplus}{a_i}$ = $a_0 \oplus a_2 \oplus \cdots \oplus a_{N-1}$. \item \textbf{Separation of the senders:} The value of a sent bit $a_i$ from the $i^{th}$ sender does not depend on the values of the other senders' input bits $\{ x_k | k \neq i \}$. To fix ideas, we may assume that each sender has access only to the value of her own input bit and not to the values of the other senders' input bits. In particular, we may assume that the senders are mutually spacelike separated at the events at which they choose their input bits, and that they do not communicate with one another subsequently. The local realism of classical physics requires that, in a classical locally nonlinear distributed evaluation, $a_i = p_i(x_i)$, where $x_i$ is the $i^{th}$ sender's input bit and $p_i$ is some deterministic one-bit Boolean function of one bit. The function $p_i$ varies, in general, from channel to channel, and from sender to sender. On the other hand, in an entanglement-assisted locally nonlinear distributed evaluation, $a_i$ may be a random bit obtained from the outcome of a quantum measurement on a qubit local to the $i^{th}$ sender and entangled with qubits local to the other senders; otherwise the structure of the communication bus is the same in the entanglement-assisted case. Importantly, the uploading process is unidirectional: the senders do not receive (download) any information from the channels, but only upload single bits using the $\oplus$ operation. \item \textbf{Local nonlinear computation by the receiver:} Olga's goal is to compute a sequence of $m$ \emph{output bits} \[ y_0, y_1, \ldots, y_{m-1} \] which are functions of the senders' input bits, using only the $m$ received bits that she receives from the $m$ channels. The overall accomplishment of the group is thus that Olga determines \begin{equation} \label{Olga y from x general} 
	\left[ \begin{array}{c}
	 y_0 \\ 
	 y_1 \\
	 \cdots \\
	 y_{m-1} 
	\end{array} \right] = F \left(\left[ \begin{array}{c}
	 x_0 \\ 
	 x_1 \\
	 \cdots \\
	 x_{N-1} 
	\end{array} \right] \right) \end{equation} Sometimes we will focus on only one of the bits $y_q$ computed by Olga, treating it as a function of the senders' input bits. The bits that Olga computes are called \emph{output bits}, and it is important to distinguish them from \emph{input bits}, \emph{sent bits} and \emph{received bits}. \end{enumerate} \end{definition}

\section{Classical restriction on the nonlinearity of locally nonlinear distributed evaluation}

In a classical world, the requirement that each sender has access only to her own input bit implies that the bit $a_i$ uploaded by the $i^{th}$ sender to the $j^{th}$ channel must be a function of her input bit, and not of the other senders' input bits. Thus, $a_i = (c_{ji} \cdot x_i) \oplus b_{ji}$ for some Boolean constants $c_{ji}$ and $b_{ji}$.  Even if an input $a_i$ is chosen ``stochastically,'' it can be treated as a deterministic function of $x_i$ in any given instance; this is tantamount to the assumption of ``local realism'' in classical physics. Olga thus receives the bit $\sum_{i}^{\oplus} a_i = \sum_{i}^{\oplus} (c_{ji} \cdot x_i) \oplus b_{ji}$ from the $j^{th}$ channel. Olga is then free to locally evaluate arbitrary nonlinear Boolean functions of these received bits, of which there are $m$. Therefore, \begin{equation} \label{Olga y from x classical last} 
y_{q} = h \left(\left(\mathbf{C_0} \cdot \left[ \begin{array}{c}
	 x_0 \\ 
	 x_1 \\
	 \cdots \\
	 x_{N-1} 
	\end{array} \right] \right) \oplus \left[ \begin{array}{c}
	 b_{00} \\ 
	 b_{01} \\
	 \cdots \\
	 b_{0(N-1)} 
	\end{array} \right], \cdots,\left( \mathbf{C_{m-1}} \cdot \left[ \begin{array}{c}
	 x_0 \\ 
	 x_1 \\
	 \cdots \\
	 x_{N-1} 
	\end{array} \right] \right) \oplus \left[ \begin{array}{c}
	 b_{(m-1)0} \\ 
	 b_{(m-1)1} \\
	 \cdots \\
	 b_{(m-1)(N-1)} 
	\end{array} \right] \right) 
\end{equation} where $h$ is an arbitrary Boolean function computed locally by Olga, the $C_j$ are constant Boolean row vectors $ \mathbf{C_j} = \left[\begin{array}{cccc} c_{j0} \ c_{j1} \ \cdots \ c_{j(N-1)} \end{array} \right]$, $b_{ij}$ is a constant Boolean matrix and $m$ is the number of channels. This leads to the following lemma.

\begin{lemma} \label{Classical Nonlinearity Bound}
\textbf{\textup{Classical bound on the nonlinearity of a locally nonlinear distributed evaluation}} In a locally nonlinear distributed evaluation with $m$ channels, Olga's output $y_q$, as given in \eqref{Olga y from x classical last}, must have order of nonlinearity $\leq m$ as a function of the senders' input bits.
\end{lemma}

\begin{proof}
The Boolean function $h$ in \eqref{Olga y from x classical last} is a function of only $m$ arguments, namely the $m$ received bits which Olga receives from the $m$ channels. Each such received bit is a linear function of the senders' input bits $x_i$. A Boolean function of $m$ arguments can have order of nonlinearity at most $m$. Therefore $h$ can, in the algebraic normal form representation, combine its arguments in at most an $m$-way \textbf{AND} statement. Since each argument of $h$ contains only terms linear in the $x_i$, distributing \textbf{AND} over \textbf{XOR} produces an expression with at most $m$-way \textbf{AND}s of the $x_i$. The reduction of this expression to algebraic normal form merely deletes conjunctions whose constant coefficients are $0$ (e.g., $0 \cdot x_1 \cdot x_2 \cdot x_5 \to 0$), and eliminates repeated instances of a single variable in conjunctions (e.g., $x_1 \cdot x_3 \cdot x_1 \to x_1 \cdot x_3$), and therefore preserves this property.
\end{proof}

\section{Preview of quantum-classical dichotomy in nonlinearity}

We will show that in an entanglement-assisted locally nonlinear distributed evaluation with $m$ channels, Olga's outputs $y_q$ can be $2^{m-1}$-nonlinear in the senders' input bits. This is an exponential enhancement in the ability of entanglement-assisted locally nonlinear distributed evaluations to evaluate highly nonlinear Boolean functions, as compared to classical locally nonlinear distributed evaluations, which are limited by \textbf{Lemma \ref{Classical Nonlinearity Bound}} to output bits that are at most $m$-nonlinear. This quantum-classical dichotomy will be the central result of this paper. 

\section{Classical impossibility of integer addition using locally nonlinear distributed evaluation}

To establish our quantum-classical dichotomy in nonlinearity, we now turn to a specific example: the problem of implementing integer addition using a locally nonlinear distributed evaluation. Contrary to appearances, the computation of the bits of the result of the integer addition of a series of bits, in contrast to the addition modulo-$2$, is a nonlinear Boolean computation. \begin{definition} \label{def:locally nonlinear integer addition} \textbf{Locally nonlinear distributed integer addition}

\noindent A locally nonlinear distributed evaluation implements binary integer addition if Olga's outputs $y_q$ are the binary digits $s_q$ the sum of the senders’ input bits, defined by \begin{equation} \label{basic sum} S = \sum x_i = \sum 2^q s_q \end{equation}
Note that $S$ is a non-negative natural number. If there are $N$ senders, note that the binary representation of the sum \eqref{basic sum} of the senders' input bits can be given recursively as \begin{eqnarray}
s_0 = S(\bmod 2) = x_0 \oplus \cdots \oplus x_{N-1} \nonumber \\
s_1 = \left(\frac{S-2^0s_0}{2^1}\right)(\bmod 2) \nonumber \\
s_2 = \left(\frac{S-2^1s_1-2^0s_0}{2^2}\right)(\bmod 2) \nonumber \\
\cdots \nonumber \\
s_q = \left(\frac{S-\sum_{p=0}^{q-1} 2^p s_p}{2^q}\right)(\bmod 2) \label{recursive sum}
\end{eqnarray} \end{definition}

It is easy to verify that all the bits $s_j$ of the integer sum, $S$, except for the least significant bit $s_0$, are nonlinear Boolean functions of the $x_i$. In particular, the bit $s_q$ is $2^q$-nonlinear, since its algebraic normal form as a function of the $x_i$ can be written as \begin{equation} \label{formula_for_binary_digits} s_q = \sum_{\oplus}{[\text{\textbf{AND} statements of length } 2^q \text{ with no repeated variables or redundant permutations}]} \end{equation} For example, for the addition of three bits $a$, $b$ and $c$ we have a two-bit output $s_0 = a \oplus b \oplus c$ and $s_1 = a \cdot b \oplus b \cdot c \oplus a \cdot c$, while for the addition of four bit $a$, $b$, $c$ and $d$ we have $s_0 = a \oplus b \oplus c \oplus d$, $s_1 = a \cdot b \oplus b \cdot c \oplus c \cdot d \oplus a \cdot c \oplus a \cdot d \oplus b \cdot d$ and $s_2 = a \cdot b \cdot c \cdot d$.
When there are N senders, the most significant bit in the binary representation of the sum of the senders' input bits is $s_L$, where $L=\lfloor \log_2(N) \rfloor$. Here, $\lfloor r \rfloor$ denotes the integer part (``floor'') of a real number $r$. If $N$ is a power of 2, then $s_L$ is just the multi-way \textbf{AND} of the senders' input bits, i.e., $s_L = x_0 \cdots x_{N-1}$. 

Thus, the order of nonlinearity of the $q^{th}$ binary digit resulting from integer addition is exponentially nonlinear in $q$. Functions with exponential nonlinearity of this sort cannot be computed efficiently using locally nonlinear distributed evaluations, as is proved in the following theorem. 
\begin{theorem} \textbf{\textup{Classical restriction on locally nonlinear distributed integer addition}}: Subject to local realism, locally nonlinear distributed integer addition is impossible when the number of channels is equal to the number of output bits.
\end{theorem}
\begin{proof}
By \textbf{Lemma \ref{Classical Nonlinearity Bound}}, Olga's output $y_q$, in a classical locally nonlinear distributed evaluation, is at most $m$-nonlinear. For integer addition with $N$ senders, there are only $\lfloor \log_2(N) \rfloor + 1$ output bits, so \begin{equation*}m = \lfloor \log_2(N) \rfloor + 1\end{equation*} Therefore, a classical locally nonlinear distributed evaluation of integer addition would---if such a protocol existed---produce output bits that are at most $\lfloor \log_2(N) \rfloor + 1$-nonlinear in the $x_i$. In contrast, correctly performing integer addition would require $y_q$ to be the function $s_q$ from \eqref{formula_for_binary_digits}, which is $2^q$-nonlinear in the $x_i$. This would exponentially violate \textbf{Lemma \ref{Classical Nonlinearity Bound}}: as $q$ increases from $0$ to $m-1 = \lfloor \log_2(N) \rfloor$, we will eventually have \begin{equation*}2^q > m = \lfloor \log_2(N) \rfloor + 1\end{equation*} so long as $N>2$, which violates \textbf{Lemma \ref{Classical Nonlinearity Bound}}. In particular, consider the last digit $s_{m-1}$ of the sum: $s_{m-1}$ is $2^{m-1}$-nonlinear in the $x_i$, and we have \begin{equation*}2^{m-1} = 2^{\lfloor \log_2(N) \rfloor} > m = \lfloor \log_2(N) \rfloor + 1\end{equation*} so long as $N>2$, with an exponentially growing violation as $N$ increases. Hence, locally nonlinear distributed implementations of integer addition are impossible in classical scenarios where the number of channels is equal to the number of output bits.
\end{proof}

\section{Quantum-classical dichotomy in nonlinearity}

We now show that an \emph{entanglement-assisted} locally nonlinear distributed evaluation \emph{can} output the binary digits $s_q$ in the sum of the senders' input bits, in contrast to the classical case. This leads to the following theorem.

\begin{theorem} \label{Q-C dichotomy} \textup{\textbf{Quantum-classical dichotomy in nonlinearity:}} an entanglement-assisted locally nonlinear distributed evaluation can evaluate Boolean functions with orders of nonlinearity exponentially greater than would be achievable in a classical locally nonlinear distributed evaluation with the same number of channels. In particular, an entanglement-assisted locally nonlinear distributed evaluation with $m$ channels can output the last binary digit $s_{m-1}$ in the sum of the sender's input bits, which is a $2^{m-1}$-nonlinear function of the senders' input bits. We have already shown, in contrast, that a locally nonlinear distributed evaluation with $m$ channels, in a classical world, can output Boolean functions with orders of nonlinearity at most $m$.
\end{theorem}

We now proceed to prove \textbf{Theorem \ref{Q-C dichotomy}}. To do so, we describe a quantum protocol for implementing integer addition via an entanglement-assisted locally nonlinear distributed evaluation, based on the recursive formulae \eqref{recursive sum} for a given digit of the sum in terms of less significant digits. 

Our protocol bears close formal relations with several results in the existing literature. Most significantly, it is an extension of the quantum pseudo-telepathy protocol known as Mermin's Parity Game \cite{RecastingMerminsMultiplayerGame}. In Mermin's Parity Game, a set of parties, indexed by a number $i$, each receive single bits $x_i$. Subject to the constraint that $\sum{x_i}$ is even, they are challenged to produce bits $y_i$ such that $\sum{y_i} = \frac{1}{2} \sum{x_i} (\bmod 2)$, \emph{without any communication occurring among the parties}. This is impossible in a classical world, but it is achievable with shared entanglement. Although the setup for locally nonlinear distributed evaluation is quite different from the symmetric setup of Mermin's Parity Game, our integer addition protocol relies crucially on repeated applications of this communication-free, entanglement-assisted operation to distributedly check the parity of one half of a distributed sum. Variants of this distributed parity check operation have been used to help derive the the maximal detector efficiency below which quantum correlations become indistinguishable from correlations achievable in local hidden variables models, and the minimal amount of superluminal classical communication which would be needed to simulate quantum correlations in local hidden variables models \cite{CombinatoricsAndQuantumNonlocality}. The same distributed parity check operation has also been used in the domain of quantum anonymous transmissions \cite{QuantumAnonymousTransmissions}.

Mermin's Parity Game is itself a generalization of Mermin's famous ``arguments without inequalities'' for quantum non-locality \cite{WhatsWrongWithTheseElementsOfReality}\cite{ExtremeQuantumEntanglementInASuperpositionOfMacroscopicallyDistinctStates}, which in turn built on the pioneering work of Greenberger, Horne and Zeilinger.  These works have given rise to an extensive literature on ``XOR games,'' such as \cite{QuantumXORProofSystems} and \cite{LimitOnNonlocalityInAnyWorldInWhichCommunicationComplexityIsNotTrivial}. Such works consider, in scenarios distinct from the present locally nonlinear distributed evaluation setup, the efficiency with which separated parties possessing local bits can generate inputs to linear Boolean operations so as to approximate nonlinear Boolean functions of their local bits. The present work can be considered as a contribution to this literature using a novel communication setup and quantum protocol.

Our protocol also bears relations with various quantum voting protocols \cite{TowardsQuantumBasedPrivacyAndVoting} \cite{QuantumProtocolsForAnonymousVotingAndSurveying} \cite{AQuantumSecretBallot}, and can be considered to be a quantum voting protocol in and of itself. This stems from the fact that only information about the sum of the senders' input bits (i.e., the ``tally'' of the senders' ``votes'') is encoded in the globally accessible classical variables and quantum states utilized in the protocol. No further information, over and above this sum, is encoded that would allow Olga or any of the senders to determine the individual values of the senders' input bits. The protocol's reliance on local phase rotations to produce global phase changes (see below) is reminiscent of the versatile quantum phase estimation technique, the basis for many quantum algorithms \cite{NielsenAndChuang}. In the phase estimation technique, this effect is used to transfer phases between two registers. The recipient register is then processed using an inverse quantum discrete Fourier transform.

\begin{definition} \label{quantum adder} \textbf{Quantum locally nonlinear distributed adder}
	
\begin{description}

\item The protocol is an entanglement-assisted locally nonlinear distributed evaluation. The scheme is identical to that of a classical locally nonlinear distributed evaluation, except that the senders choose their sent bits on the basis of local measurements on the constituent qubits of GHZ states shared among the senders and Olga. Olga computes the output bits given the received bits using local qubit rotations and measurements in addition to classical logic. This is in contrast to what occurs in a classical locally nonlinear distributed evaluation, wherein each sender chooses her sent bits to be deterministic functions of her input bit, and wherein Olga computes the output bits given the received bits using only classical logic.

\item As introduced before, there are N senders and one receiver, Olga. The $i^{th}$ sender possesses an input bit $x_i$.

\item \textbf{Entanglement resource:} $\lfloor \log_2(N) \rfloor$ copies of a $(N+1)$-qubit GHZ state: $\ket{0}^{\otimes (N+1)} + \ket{1}^{\otimes (N+1)}$. One qubit from each copy of the GHZ state is given to each sender, and one qubit from each copy of the GHZ state is given to Olga.

\item \textbf{Communication resource:} There are $m = \lfloor \log_2(N) \rfloor + 1$ one-bit channels of the type shown in \textbf{Figure \ref{fig:Figure-1}}. A non-negative integer $q$ indexes channels. Note that there are as many channels as there are bits in the integer sum, $S$, but that one fewer GHZ state is used.
	
\item[1] \textbf{Preparation and uploading of the sent bits:} Each sender uploads her input bit $x_i$ to channel $q=0$. The received bit from channel $q=0$ is $x_0 \oplus x_1 \oplus \cdots \oplus x_{N-1}$. Note that no GHZ state is consumed during the use of channel $q=0$. 

The procedure for the other channels is less direct. For each channel $q=1,\cdots,m-1$, a separate GHZ state is used, and all senders obey the following protocol.

\begin{description}

\item[a] \textbf{Local rotation by senders:} The $i^{th}$ sender rotates her qubit in the $q^{th}$ GHZ state about the z-axis by an angle \begin{equation*} -\pi \cdot \frac{x_i}{2^q}\end{equation*} so that the global state becomes \begin{equation*}\ket{0}^{\otimes (N+1)} + e^{i\frac{\pi}{2^q} \sum x_l} \ket{1}^{\otimes(N+1)}\end{equation*} Note that in the last expression, the index $l$ is a dummy variable indexing the $N$ senders.

\item[b] \textbf{Local Hadamard by senders:} The $i^{th}$ sender applies the $H$ (\textbf{Hadamard}) gate to her qubit, so that the global state becomes \begin{eqnarray}(\ket{0}+\ket{1})^{\otimes N} \otimes \ket{0}_{Olga} + e^{i\frac{\pi}{2^q} \sum x_l} (\ket{0}-\ket{1})^{\otimes N} \otimes \ket{1}_{Olga} \nonumber \\ = (\sum_j{\ket{e_{Nj}}} + \sum_j{\ket{o_{Nj}}}) \otimes \ket{0}_{Olga} + e^{i\frac{\pi}{2^q} \sum x_l}(\sum_j{\ket{e_{Nj}}} - \sum_j{\ket{o_{Nj}}}) \otimes \ket{1}_{Olga}\end{eqnarray} Here $\sum_j \ket{e_{Nj}}$ is the sum over all N-qubit computational basis states with even Hamming weight, and $\sum_j\ket{o_{Nj}}$ is the sum over all N-qubit computational basis states with odd Hamming weight.

\item[c] \textbf{Local measurement and uploading by senders:} Each sender measures her qubit in the computational basis (z-direction) to obtain a result $0$ or $1$. These measurement results become the sent bits uploaded to channel $q$. It is useful to distinguish two cases. In \textbf{Case 1}, the senders' measurements project the subspace spanned by the states of the senders' qubits into a computational basis state of even Hamming weight, Olga's received bit from channel $q$ is $0$, and the global state becomes \begin{equation*} \ket{e_{Nk}} \otimes (\ket{0}_{Olga} + e^{i\frac{\pi}{2^q} \sum x_l} \ket{1}_{Olga})\end{equation*} for some $k$. In \textbf{Case 2}, the senders' measurements project the subspace spanned by the states of the senders' qubits into a computational basis state of odd Hamming weight, Olga's received bit from channel $q$ is $1$ and the global state becomes \begin{equation*} \ket{o_{Nk}} \otimes (\ket{0}_{Olga} - e^{i\frac{\pi}{2^q} \sum x_l}\ket{1}_{Olga})\end{equation*} for some $k$.

\end{description}

\item[2] \textbf{Olga's computation of the output bits:} Olga obtains the received bits from each of the $m$ channels. For channel $q=0$, the senders have merely uploaded their input bits to the channel, the received bit is $x_0 \oplus x_1 \oplus \cdots \oplus x_{N-1}$, and Olga simply produces this directly as her output bit $y_0$. Note that $y_0 = s_0$, the desired least significant bit in the sum of the senders' input bits. Then, addressing each of the channels $q=1,\cdots,m-1$ in order, Olga performs the following operations using her qubit from the corresponding GHZ state.

\begin{description}

\item[a] \textbf{Local rotation by Olga:} Olga performs a rotation of her qubit about the z-axis by an angle \begin{equation*}\alpha_q \equiv -\pi \cdot \frac{ \sum_{p=0}^{q-1} 2^p y_p }{2^q}\end{equation*} which she computes from her store of \emph{previous} output bits. In \textbf{Case 1}, the global state becomes \begin{equation*} \ket{e_{Nk}} \otimes (\ket{0}_{Olga} + e^{i(\alpha_q + \frac{\pi}{2^q} \sum x_l)}\ket{1}_{Olga}) \end{equation*} whereas in \textbf{Case 2} the global state becomes \begin{equation*} \ket{o_{Nk}} \otimes (\ket{0}_{Olga} - e^{i(\alpha_q + \frac{\pi}{2^q} \sum x_l)}\ket{1}_{Olga}) \end{equation*} Inspecting the expression $e^{i(\alpha_q + \frac{\pi}{2^q} \sum x_l)}$, and comparing with \eqref{recursive sum}, we see that \begin{equation*}e^{i(\alpha_q + \frac{\pi}{2^q} \sum x_l)} = e^{i \pi s_q} = \pm 1\end{equation*} Using this simplification, we see that in \textbf{Case 1}, the global state becomes \begin{equation*} \ket{e_{Nk}} \otimes (\ket{0}_{Olga} + e^{i \pi s_q} \ket{1}_{Olga}) \end{equation*} whereas in \textbf{Case 2} the global state becomes \begin{equation*}\ket{o_{Nk}} \otimes (\ket{0}_{Olga} - e^{i \pi s_q}\ket{1}_{Olga}) \end{equation*}

\item[b] \textbf{Local Hadamard by Olga:} Olga performs the $H$ operation on her qubit. 

\begin{enumerate} 
	
\item If $s_q = 0$, then in \textbf{Case 1} the global state becomes $\ket{e_{Nk}} \otimes \ket{0}_{Olga}$ and in \textbf{Case 2} the global state becomes $\ket{o_{Nk}} \otimes \ket{1}_{Olga}$. 

\item If $s_q = 1$, then in \textbf{Case 1} the global state becomes $\ket{e_{Nk}} \otimes \ket{1}_{Olga}$ and in \textbf{Case 2} the global state becomes $\ket{o_{Nk}} \otimes \ket{0}_{Olga}$.
\end{enumerate}

\item[c] \textbf{Local measurement and production of the $q^{th}$ output bit by Olga:} Olga measures her qubit in the computational basis (z-direction), obtaining a result $0$ or $1$. She computes the \textbf{XOR} of the measurement result with the received bit from channel $q$ and records this as the output bit $y_q$. 
\begin{enumerate}
\item If $s_q = 0$, then in \textbf{Case 1} the received bit is $0$ and Olga's measurement result is $0$ and in \textbf{Case 2} the received bit is $1$ and Olga's measurement result is $1$. In either case $y_q = 0 = 0 \oplus 0 = 1 \oplus 1 = s_q$. 

\item If $s_q = 1$, then in \textbf{Case 1} the received bit is $0$ and Olga's measurement result is $1$ and in \textbf{Case 2} the received bit is $1$ and Olga's measurement result is $0$. In either case $y_q = 1 = 0 \oplus 1 = 1 \oplus 0 = s_q$. 
\end{enumerate}
Therefore Olga's output bit $y_q$ faithfully reproduces the desired digit $s_q$ of the sum of the senders' input bits, and hence this protocol implements a locally nonlinear distributed integer addition!

\end{description}

\end{description}

\end{definition}

In the above protocol, the senders first upload to all $m$ channels before Olga begins performing quantum operations. In a second stage, the received bits from all $m$ channels are used by Olga to locally compute the output bits. Alternatively, an iterative entanglement-assisted protocol exists which accomplishes the same function using the same communication constraints. In the iterative version, both the senders and Olga perform local qubit rotations, Hadamards and measurements during each iteration. Only one channel and at most one GHZ state are used per iteration (the first iteration requires no GHZ state). The iterative version of the protocol produces one correct output bit per iteration and leads to a somewhat simpler formal description. The iterative and non-iterative versions are essentially equivalent descriptions of a single protocol, due to symmetry with respect to the time-ordering of operations that are performed at spacelike separation. We chose to use the above non-iterative description here for ease of direct comparison with our description of the general locally nonlinear distributed evaluation scheme, in which there is no natural notion of iterative cycles.

\section{Conclusion}
\label{sec:Summary and Discussion}

We have described an entanglement-assisted protocol for integer addition which operates subject to the same communication handicap as in a classical locally nonlinear distributed evaluation. The output bits $y_q = s_q$ in the entanglement-assisted protocol can be $2^{m-1}$-nonlinear Boolean function of the senders' input bits, where $m$ is the number of channels, whereas any classical locally nonlinear distributed evaluation can output bits that are at most $m$-nonlinear Boolean functions of the senders' input bits. We have thus demonstrated an exponential enhancement in the efficiency with which nonlinear Boolean functions can be computed in locally nonlinear distributed evaluation protocols that are given access to local rotations and measurements on maximally-entangled multi-partite quantum states. The quantum protocol for locally nonlinear distributed addition with $N$ senders requires fine, non-Clifford group \cite{GeometricApproachToDigitalQuantumInformation}, local controlled state vector rotations on the constituent qubits of a set of $(N+1)$-qubit GHZ states. A question for further research is whether, in more general distributed computations, fine local controlled rotations on entangled qubits can enhance the computational power of communication-sparse systems with local nonlinear logic linked by linear communication channels. Can exponential or polynomial enhancement be obtained for other functions? What determines the enhancement in general?

\section{Acknowledgements}
\label{sec:Acknowledgements}
AHM was supported by the Yale College Perspectives on Science summer research fellowship in summer 2006. MD acknowledges partial support by Coll\`ege de France.
\bibliographystyle{plain}
\bibliography{Devoret_Marblestone_DRAFT12-revised}
\end{document}